\documentclass[12pt]{amsart}
\usepackage[english]{babel}
\usepackage{amsmath}
\usepackage{amsthm}
\usepackage{amssymb}
\usepackage{mathrsfs}
\usepackage{enumerate}
\usepackage{physics}
\usepackage[notcite, final, notref]{showkeys}
\usepackage{dsfont}
\usepackage[dvips]{color}
\newtheorem{theorem}{Theorem}[section]

\newtheorem{lemma}[theorem]{Lemma}
\newtheorem{proposition}[theorem]{Proposition}

\newtheorem{definition}[theorem]{Definition}

\theoremstyle{definition}
\newtheorem{remark}[theorem]{Remark}

\newtheorem{example}[theorem]{Example}

\newcommand{\ov}{\overline}

\newcommand{\w}{\omega}

\newcommand{\BC}{\mathbb{C}}
\usepackage{xcolor}

\title[Commutators on Fock spaces]{Commutators on Fock spaces}

\author[D. Alpay]{Daniel Alpay}
\address{(DA)
Faculty of Mathematics, Physics, and Computation\\
Schmid College of Science and Technology\\
Chapman University\\
One University Drive
Orange, California 92866\\
USA}
\email{alpay@chapman.edu}

\author[P. Cerejeiras]{Paula Cerejeiras}
\address{(PC) CIDMA - Center for Research and Development in Mathematics and Applications, \newline Department of Mathematics, University of Aveiro \newline Campus Universit\'ario de Santiago
\newline 3810-193 Aveiro, Portugal}
\email{pceres@ua.pt}

\author[U. K\"ahler]{Uwe K\"ahler}
\address{(UK) CIDMA - Center for Research and Development in Mathematics and Applications, \newline Department of Mathematics, University of Aveiro
\newline Campus Universit\'ario de Santiago \newline 3810-193 Aveiro, Portugal}
\email{ukaehler@ua.pt}

\author[T. Kling]{Trevor Kling}
\address{(TK) Faculty of Mathematics, Physics, and Computation\\
Schmid College of Science and Technology\\
Chapman University\\
One University Drive
Orange, California 92866\\
USA\\ {Present address:}
Department of Physics and Astronomy\\ Purdue University\\ West Lafayette, Indiana 47907, USA
}
\email{klingt@purdue.edu}

\begin{document}

\begin{abstract}
  Given a weighted $\ell^2$ space with weights associated to an entire function, we consider pairs of weighted shift operators, whose commutators are diagonal operators, when considered as operators over a general Fock space. We establish a calculus for the algebra of these commutators and apply it to the general case of Gelfond-Leontiev derivatives. This general class of operators includes many known examples, like classic fractional derivatives and Dunkl operators. This allows us to establish a general framework which goes beyond the classic Weyl-Heisenberg algebra. Concrete examples for its application are provided. 

\end{abstract}

\maketitle


\section{Introduction}
\setcounter{equation}{0}
It is a well-known fact that the Bargmann-Fock space arises via the Bargmann representation of the Heisenberg group. This results in a close connection between these two; see \cite{MR2934601}, \cite{Folland} or \cite{Perelomov}. Moreover, this fact also explains why the classic Bargmann-Fock space is a central object in quantum physics. The importance of the Bargmann-Fock space stems from the fact that in this space the dual of the derivative operator is the multiplication operator $M_z.$ Since these operators are Fourier (as well as Fischer) duals of each other it provides the basis for the study of pseudo-differential operators on Fock spaces as well as of Toeplitz operators. Furthermore, the Lie group induced by the arising Lie algebra has some interesting consequences such as its translation invariance.

This leads automatically to the question if we can use the same or similar algebraic methods in other settings where the duals with respect to integral transform are different from the above.  Instead of concentrating on the derivative and multiplication operators whose commutator is the identity and, therefore, leads immediately to the Heisenberg group~\cite{Folland}, we first look at the dual pair of integration and back-shift operators. However, the commutator between the two is not the identity operator, but rather a (infinite-dimensional) diagonal operator when applied to the standard (non-normalized) basis $e_n(z)=z^n$. This is an important point since in many other cases, like fractional derivatives of Gelfond-Leontiev type, we encounter similar algebraic properties. In fact, as will be seen later the class of Gelfond-Leontiev operators of generalized differentiation provides a general setting which includes standard examples like classic fractional derivatives (such as Caputo and
Riemann-Liouville derivatives, the latter by changing the ground state~\cite{Kiryakova}) and Dunkl (or difference-differential) operators, and, therefore, has a broad range of applications in physics. We only recall that while fractional differential operators are being applied in many areas, like fractional mechanics, Dunkl operators appear naturally in the study of Calogero-Sutherland-Moser models for $n$-particle systems ~\cite{Rostler2003}. In these cases even the commutator of the derivative and multiplication operator $M_z$ does not lead to the identity operator, but to a diagonal operator when applied to the standard basis. While this leads to the well-known fact that in general there is no finite Leibniz rule for fractional derivatives it also means that the resulting Lie algebra is
much richer. This means that we need an approach which allows us to work with such structures where the commutator is a diagonal operator when applied to an appropriate basis. To this end we are going to study commutator relations over generalized Fock spaces. As principal example we are going to look into the important case of the backward-shift operator which is dual of
the integration operator in the standard case. This allows us to consider settings which a priori are quite different, but which fit in the general analysis we are doing.\smallskip

The outline of the paper is as follows.
In Section \ref{sec-21} we discuss the backward shift operator in the classical Fock space.
In Section \ref{Sec2} we study the Fock space associated to the Gelfond-Leontiev operator. A first example of the calculus on diagonal is given in Section \ref{Sec3}, where the family of spaces  studied in \cite{MR3816055} is considered. Finally,
in Section \ref{Sec4} we consider the general setting of pairs of weighted shift operators.

\section{The backward shift operator in the Fock space}
\label{sec-21}
By starting with the backward-shift operator we can calculate its dual in Fock spaces with more general weights than usually considered in the literature, but which continue to be reproducing kernel Hilbert spaces. \smallskip

Let us recall the definition of the Fock space. Let $dA(z) := \frac{1}{\pi}e^{-|z|^2} dxdy, ~z=x+iy,$ be a weighted Lebesgue measure in $\mathbb{C}.$ The classic Fock space $\mathcal{F}_1$
is the set of all entire functions $f : \mathbb{C} \mapsto \mathbb{C}$ such that
\[
\| f\|_{1} := \left(  \frac{1}{\pi}\int_{\mathbb{C}} |f(z)|^2 dA(z) \right)^{\frac{1}{2}} =\sum_{n=0}^\infty n!|f_n|^2 < \infty,
\]
where $f(z)=\sum_{n=0}^\infty f_nz^n$.

The classical Fock space $\mathcal F_1$ can be seen as the unique (up to a strictly positive multiplicative factor for the norm) Hilbert space of power series defined in a neighborhood of the origin for which the complex derivative $\partial$ and the operator $M_z$ of multiplication by the variable $z$ are closed operators defined on the span of polynomials and adjoint to each other there. Their commutator
\begin{equation}
  [\partial, M_z]=\partial M_z-M_z\partial= \mathbf{id}
\end{equation}
is equal to the identity operator $\mathbf{id}$. In fact it then follows that the elements are entire functions, see \cite{MR0157250,bargmann}. \\


Another important operator in this context is the backward-shift operator $R_0$ defined by
\begin{equation}
  (R_0f)(z)=\begin{cases} \,\dfrac{f(z)-f(0)}{z},\quad z\not=0,\\
    \, f^\prime(0),\quad\hspace{13mm} z=0,\end{cases}
  \end{equation}
  for functions analytic in a neighborhood of the origin. It is easily seen (see  lemma below) that $R_0$ is a contraction from the Fock space into itself, with its adjoint being the integration
  operator $I$ defined by
\begin{equation} \label{IntOp}
  (If)(z)=\int_{[0,z]}f(s)ds
\end{equation}
where the functions are assumed analytic in an open convex neighborhood of the origin (and in the whole of $\mathbb C$ in the case of the Fock space).

\begin{lemma} \label{partial-int}
  For the case of the classic Fock space $\mathcal F_1$ we have for the backward-shift operator $R_0$ the following properties:
  \begin{enumerate}[(1)]
  \item $R_0$ is a contraction in the Fock space.
\item Its adjoint operator $R_0^\ast$ is the integration operator.
\end{enumerate}
\end{lemma}
\begin{proof}
  Let $f\in\mathcal F_1,$ that is, $f$ is an entire function $f(z)=\sum_{n=0}^\infty f_nz^n$ with finite $\mathcal F_1$-norm $$\| f \|_1^2 = \sum_{n=0}^\infty |f_n|^2n!<\infty.$$ We have
  \[
(R_0f)(z)=\sum_{n=0}^\infty f_{n+1}z^n,
\]
and so
\[
\|R_0f\|_{1}^2=\sum_{n=0}^\infty|f_{n+1}|^2n!\le\sum_{n=0}^\infty|f_{n+1}|^2(n+1)!\le\|f\|_{1}^2.
\]
Since $R_0$ is bounded, it is enough to check that $R_0^*=I$ on monomials. Let $n,m\in\mathbb N_0$. We have
\[
  \langle I z^n, z^m\rangle_{1}  =\left\langle \frac{{z^{n+1}}}{n+1}, z^m\right\rangle_{1}=\delta_{n+1,m}\frac{(n+1)!}{n+1}=\delta_{n+1,m}n!,
  \]
  while, for $m\ge 1$,
  \[
    \langle {z^n},R_0 z^m\rangle_{1}=\langle {z^n}, z^{m-1}\rangle_{1}=\delta_{n,m-1}n!,
  \]
  and hence the result for $m\ge 1$. For $m=0$, the equality is trivial:
  \[
    \langle I {z^n},1\rangle_{1}=\left\langle\frac{{z^{n+1}}}{n+1},1\right\rangle_{1}=0\quad {\rm while}\quad \langle {z^n},R_0 1\rangle_{1}=\langle {z^n},0\rangle_{1}=0.
    \]
\end{proof}

The condition $R_0^*=I$ in fact is a characterization of the Fock space (up to a strictly positive multiplicative factor for the norm). Now the commutator of $R_0$ and $I$ is equal on monomials to
\begin{equation}
  \label{commutator-R0I}
  [R_0, I]({z^n})= \begin{cases} \,\, 1,\hspace{16.2mm}\,\,\, n=0,\\
  			  \,\,-\dfrac{{z^n}}{n(n+1)},\,\,n=1,2,\ldots
   			 \end{cases}.
\end{equation}
      We will denote by $D_0$ the (formal) diagonal operator defined by the right hand side of \eqref{commutator-R0I}, i.e.
\begin{eqnarray}
  D_0 & = & \left( \begin{array}{cccccc}
  1 & 0 & 0 & \cdots & 0 & \cdots \\
 0 & -\frac{1}{2} & 0 & \cdots & 0 & \cdots \\
 0 & 0 & -\frac{1}{2\cdot 3} & \cdots & 0 & \cdots  \\
 \vdots &  \vdots &  \vdots & ~ & \vdots & ~\\
 0 & 0 & 0 & \cdots & -\frac{1}{n(n+1)} & \cdots \\
  \vdots &  \vdots &  \vdots & ~ & \vdots & ~
  \end {array} \right) \nonumber \\
      & := & {\rm diag}\, \left(1,-\frac{1}{2},-\frac{1}{2\cdot 3},\ldots, -\frac{1}{n(n+1)},\ldots\right),
             \label{d0}
\end{eqnarray}
and so
\begin{equation}
  [R_0,I]=D_0.
  \label{d0d0}
  \end{equation}
It is interesting to already note that (still formally at this stage)
\begin{equation}
{\rm Tr}\, D_0=0.
\end{equation}

We remark that formula \eqref{commutator-R0I} makes sense also for such spaces different from the Fock space in which case we do not have $R_0^*=I.$ Furthermore, \eqref{commutator-R0I} suggests that the study of the algebra generated by $R_0$ and $I$ requires also diagonal operators as coefficients.\\


%

%

\section{Fock spaces related to fractional derivatives}
\setcounter{equation}{0}
\label{Sec2}

We begin with the definition of generalized differentiation and integration operators with respect to a given entire function. Let the function
\begin{equation} \label{Eq:EntireFunction}
\varphi(z) =\sum_{n=0}^\infty  \frac{z^n}{\varphi_n},
\end{equation}
be an entire function with order $\rho >0$ and  degree $\sigma > 0,$ that is, such that $\lim_{n \rightarrow \infty} \frac{n^{\frac{1}{\rho}} }{\sqrt[n]{|\varphi_n |}} =\left(\sigma e \rho\right)^{\frac{1}{\rho}}$.\\

Furthermore we always assume the normalization
\begin{equation}
  \varphi_0=1.
\end{equation}
\begin{definition} Assume that $\varphi_n>0$ for $n\in\mathbb N_0$.
  \label{Hvarphi} We define the Hilbert space $\mathcal H(\varphi)$ as the set of all entire functions endowed with the inner product
  \begin{equation}
    \label{krein-krein}
\langle f,g \rangle_\varphi=\sum_{n=0}^\infty  \varphi_n  f_n \ov g_n,
    \end{equation} where $f(z)=\sum_{n=0}^\infty f_n z^n$ and  $g(z)=\sum_{n=0}^\infty g_n z^n.$
\end{definition}

The inner product \eqref{krein-krein} induces the norm
  \begin{equation}
    \|f\|^2_\varphi=\sum_{n=0}^\infty \varphi_n  |f_n|^2, \quad f \in \mathcal H(\varphi).
    \end{equation}

\begin{proposition} The Hilbert space
  $\mathcal H(\varphi)$ is a reproducing kernel Hilbert space with reproducing kernel
  \begin{equation}
k(z,\w)=\varphi(z\overline{\omega}).
    \end{equation}
  \end{proposition}

    \begin{remark}{\rm
When the sequence $\varphi_n$ is increasing and $\varphi_0=1$ the space $\mathcal H(\varphi)$ is a  de Branges Rovnyak space. See \cite{MR4032209}.}
      \end{remark}

\begin{definition}
Let $\varphi$ be as in (\ref{Eq:EntireFunction}). We define the Gelfond-Leontiev (G-L) operator of generalized differentiation with respect to $\varphi,$ denoted as $\partial^\varphi,$
as the operator acting on function $f$ analytic in a neighborhood of the origin, $f(z) =\sum_{n=0}^\infty f_n z^n$,  by:
\begin{equation} \label{Eq:p1+p2_dif}
f(z) =\sum_{n=0}^\infty f_n z^n \quad \mapsto \quad \partial^\varphi f(z) =\sum_{n=1}^\infty f_n \frac{\varphi_{n}}{\varphi_{n-1}} ~z^{n-1}. 
\end{equation}
\end{definition}

It is well known that $\left(\partial^\varphi\right)^*=M_z$ over the space $\mathcal H(\varphi)$. We now introduce the corresponding generalization of the integration operator $I$:

\begin{proposition}
The operator $R_0$ is densely defined and closed in $\mathcal H(\varphi)$. It is a contraction when the sequence $(\varphi_n)_{n=0}^\infty$ is non-decreasing.  The adjoint of $R_0$ is given by
\begin{equation}
  I^\varphi z^n=\frac{\varphi_n}{\varphi_{n+1}}z^{n+1}, \quad n=0, 1, 2, \ldots.
  \end{equation}
\end{proposition}

\begin{proof} The operator $R_0$ is closed due to the fact that in a reproducing kernel Hilbert space, convergence in norm implies pointwise convergence. We consider a sequence of 
functions $\tilde f_k : \mathbb{C} \mapsto \mathbb{C}$ in  $\mathcal H(\varphi)$ converging in norm to $f\in\mathcal H(\varphi)$, and such that the sequence $(R_0\tilde f_k)_{k=1}^\infty$ is convergent, with limit $g\in\mathcal H(\varphi)$. Then for every $\w\not=0$ in the domain of
  analyticity of the elements of $\mathcal H(\varphi)$
  \[
    \begin{split}
      \lim_{k\rightarrow\infty}\tilde f_k(\w)&=f(\w)\\
       \lim_{k\rightarrow\infty} R_0 \tilde f_k (\w)=\lim_{k\rightarrow\infty}\frac{\tilde f_k(\w)-\tilde f_k(0)}{\w}&=g(\w).
        \end{split}
  \]
  It follows that $g(\w)=R_0f(\w)$ for $\w\not=0$ and for $\w=0$ by analytic continuation.\smallskip

Since the sequence $(\varphi_n)_{n=0}^\infty$ is non-decreasing, that is  $\varphi_n\le\varphi_{n+1}$ for $n\in\mathbb N_0$, we have
$$R_0 f(z) = R_0\left( \sum_{n=0} f_n z^n  \right) = \sum_{n=1} f_{n} z^{n-1},$$
and hence
  \begin{equation}
    \label{structure}
   \| R_0 f \|_\varphi^2=\sum_{n=1}^\infty\varphi_{n-1}|f_n|^2\le\sum_{n=1}^\infty\varphi_{n}|f_n |^2\le\|f\|^2_\varphi-|f_0|^2.
\end{equation}
We now compute the adjoint of $R_0$. Let $n,m\in\mathbb N_0$. We have:
  \[
    \begin{split}
      \langle R_0 z^n, z^m\rangle_{\varphi}&=     \begin{cases} \langle z^{n-1},z^m\rangle_{\varphi} = \varphi_{n-1}\delta_{n-1,m}, \quad n\ge 1,\\
        0,\quad \hspace{17.5mm}n=0,\end{cases}
    \end{split}
    \]
      while
      \[ \langle z^n,I^\varphi z^m\rangle_{\varphi}=      \langle z^n, \frac{\varphi_m}{\varphi_{m+1}} z^m\rangle_{\varphi}  = \varphi_{m+1}\frac{\varphi_m}{\varphi_{m+1}}\delta_{n,m+1}= \varphi_{m}\delta_{n,m+1} =    \begin{cases}\varphi_{n-1}\delta_{n-1,m}, \quad n\ge 1,\\
        0,\quad \hspace{17mm}n=0,\end{cases},
   \]
    so that $R_0^*=I^\varphi$.
    \end{proof}

We note that inequality \eqref{structure} is the structure inequality which characterizes de Branges Rovnyak spaces; see \cite[Theorem 3.1.2, p. 83]{adrs} and \cite{MR4032209} for further on this point.
\begin{definition}
  Let $\varphi$ be as in (\ref{Eq:EntireFunction}). We define the Gelfond-Leontiev (G-L) operator of generalized integration with respect to $\varphi,$ denoted as $I^\varphi,$
as the operator acting on functions $f$ analytic in a neighborhood of the origin, by 
\begin{equation} \label{Eq:p1+p2_int}
  f(z) =\sum_{n=0}^\infty f_n z^n \quad \mapsto \quad I^\varphi f(z) =\sum_{n=0}^\infty f_n \frac{\varphi_{n}}{\varphi_{n+1}} ~z^{n+1}.
\end{equation}
\end{definition}

\begin{example}\label{exML1} For $\varphi(z) = e^z$ we have $\varphi_n = \Gamma(n+1), n=0,1, 2, \ldots,$ so that
\begin{alignat}{4}
  \partial^\varphi f(z)  &= \partial^\varphi \left( \sum_{n=0}^\infty f_n z^n \right) &= \sum_{n=1}^\infty f_n ~n  z^{n-1} &= \partial_z f(z)\\
  I^\varphi f(z)  &= I^\varphi \left( \sum_{n=0}^\infty f_n z^n \right) &= \sum_{n=0}^\infty \frac{f_n}{n+1} z^{n+1} &= I f(z).
  \end{alignat}

\end{example}

\begin{example}\label{exML2}When $\varphi$ is the Mittag-Leffler function
\begin{equation}\label{Eq:ML_function}
\varphi(z) = E_{\frac{1}{\rho},\mu}(z) =\sum_{n=0}^\infty \frac{z^n}{\Gamma\left(\mu +\frac{n}{\rho}\right)}, \qquad \rho>0, ~\mu \in \BC ~s.t.~ \mathrm{Re}(\mu)>0,
\end{equation}
we have that $\varphi(z) = E_{\frac{1}{\rho},\mu}(z) $ is an entire function of order $\rho$ and type $1$ (see \cite[p. 56]{MR4179587}). We obtain $\varphi_n  =\Gamma\left(\mu
+\frac{n}{\rho}\right)$ and the operator (\ref{Eq:p1+p2_dif}) becomes the Dzrbashjan-Gelfond-Leontiev operator:
$$\partial^{\rho,\mu} f(z) =\sum_{n=1}^\infty f_n \frac{\Gamma\left(\mu +\frac{n}{\rho}\right)}{\Gamma\left(\mu +\frac{n-1}{\rho}\right)} ~z^{n-1}.$$
In a similar way, the operator (\ref{Eq:p1+p2_int}) is then
$$I^{\rho,\mu} f(z) =\sum_{n=0}^\infty f_n \frac{\Gamma\left(\mu +\frac{n}{\rho}\right)}{\Gamma\left(\mu +\frac{n+1}{\rho}\right)} ~z^{n+1}.$$

\begin{lemma}
The function $E_{\frac{1}{\rho},\mu}(z\overline{\w})$ is positive definite
if $\mu>0$.
\end{lemma}

This is an immediate consequence since for $\rho, \mu >0$ we have $\Gamma (\mu +\frac{n}{\rho}) >0$ and, therefore, $\varphi_n>0$ too.

\end{example}

\begin{example} \label{Example:3.10} A second example from a different area than the one above is the rank-one case of a Dunkl operator (also differential-difference operator) where the reflection group $\mathcal{G} = \{ \mathbf{id}, \sigma \}$ acts on $\mathbb{R}$ by the reflection $\sigma(x) = -x.$ Given a multiplicity constant $\kappa \in \mathbb{C} ~(\mbox{\rm Re}(\kappa) >0)$ we get the first-order rational Dunkl operator attached to $\mathcal{G}$ and $\kappa$ defined as
$$Tf(x) := f'(x) + \kappa ~\frac{ f(x) - f(-x) }{x}.$$
The above differential-difference operator is linked to the 
function $\varphi$ with
$$
\varphi(z)=e^{z} {}_1F_1(\kappa,2\kappa+1;-2z).
$$
We obtain as coefficients (see \cite{Rostler2003})
\begin{equation}
  \label{tyuio}
  \varphi_{2n}= \frac{(2n)!\left(\kappa+\frac{1}{2}\right)_n}{\left(\frac{1}{2}\right)_n}\quad
  \mbox{ and }\quad  \varphi_{2n+1}=\frac{(2n+1)!\left(\kappa+\frac{1}{2}\right)_{n+1}}{\left(\frac{1}{2}\right)_{n+1}},
\end{equation} where $(x)_n := x(x+1) \cdots (x+n-1), ~x \in \mathbb{R}\setminus \mathbb{Z},$ denotes the rising factorial. 
Hence, easy calculations carry
\begin{gather}
\partial^{\varphi} f(z) = \sum_{n=1}^\infty f_n \frac{\varphi_n}{\varphi_{n-1}} z^{n-1} = \sum_{n=0}^\infty f_{2n+1} \frac{\varphi_{2n+1}}{\varphi_{2n}} z^{2n} + \sum_{n=1}^\infty f_{2n} \frac{\varphi_{2n}}{\varphi_{2n-1}} z^{2n-1} \nonumber \\
= \sum_{n=0}^\infty f_{2n+1} \frac{ (2n+1)!\left(\kappa+\frac{1}{2}\right)_{n+1}  \left(\frac{1}{2}\right)_{n}}{\left(\frac{1}{2}\right)_{n+1}  (2n)! \left(\kappa+\frac{1}{2}\right)_{n}  } z^{2n}  + \sum_{n=1}^\infty f_{2n} \frac{ (2n)! \left(\kappa+\frac{1}{2}\right)_{n} \left(\frac{1}{2}\right)_n   }{ \left(\frac{1}{2}\right)_n (2n-1)! \left(\kappa+\frac{1}{2}\right)_n } z^{2n-1} \nonumber\\
= \sum_{n=0}^\infty f_{2n+1}  (2n+2 \kappa + 1) z^{2n} + \sum_{n=1}^\infty f_{2n} 2n z^{2n-1} =: Tf(z). \label{3.14}
\end{gather}
As such the rank-one operator $T$ can be seen as a special case of the Gelfond-Leontiev operator $\partial^\varphi$. Moreover, the function $\varphi(z\overline{\w})$ is positive definite since the coefficients
\eqref{tyuio} are positive.
Likewise, we obtain
\begin{gather}
I^{\varphi} f(z) = \sum_{n=0}^\infty f_n \frac{\varphi_n}{\varphi_{n+1}} z^{n+1} = \sum_{n=0}^\infty f_{2n} \frac{\varphi_{2n}}{\varphi_{2n+1}} z^{2n+1} + \sum_{n=1}^\infty f_{2n-1} \frac{\varphi_{2n-1}}{\varphi_{2n}} z^{2n} \nonumber \\
= \sum_{n=0}^\infty f_{2n} \frac{(2n)! \left(\kappa+\frac{1}{2}\right)_n  \left(\frac{1}{2}\right)_{n+1}}{\left(\frac{1}{2}\right)_n  (2n+1)!\left(\kappa+\frac{1}{2}\right)_{n+1}  } z^{2n+1}  + \sum_{n=1}^\infty f_{2n-1} \frac{ (2n-1)! \left(\kappa+\frac{1}{2}\right)_{n} \left(\frac{1}{2}\right)_n }{ \left(\frac{1}{2}\right)_n (2n)! \left(\kappa+\frac{1}{2}\right)_n } z^{2n} \nonumber\\
= \sum_{n=0}^\infty f_{2n}  \frac{z^{2n+1}}{2n+2 \kappa + 1 }  + \sum_{n=1}^\infty f_{2n-1} \frac{z^{2n}}{2n}. \label{3.15}
\end{gather}

\end{example}

\section{Diagonal operators over the spaces $\mathcal F_p$}
\setcounter{equation}{0}

\label{Sec3}
Let us start our discussion of the properties and expressions of the diagonal operators arising from the commutators with the special case of the $p$-Fock space. These spaces have been investigated \cite{MR3816055}. For us they are particularly interesting since they correspond to the choice of
$$\varphi(z) = \sum_{n=0}^\infty \frac{z^n}{(n!)^p},$$ that is, $\varphi_n=(n!)^p$ for $p \in \mathbb{N}$ or, with other words, they represent the case of powers of the factorials.

\begin{definition} We define the $p$-Fock space $\mathcal{F}_p, ~p \in \mathbb{N},$ as the set of all entire functions $f : \mathbb{C} \mapsto \mathbb{C}$ such that
$$\| f \|_p := \sum_{n=0}^\infty |f_n|^2 (n!)^p < \infty.$$
\end{definition}

Hereby, the particular case $p=1$ is the classic Fock space $\mathcal F_1$ already mentioned in Section 1.
The inner product between two functions $f, g \in \mathcal{F}_p$ is given by
\[
    \langle f, g \rangle = \sum_{n=0} f_n \overline{g_n}(n!)^p, \text{ where } f(z) = \sum_{n=0}^\infty f_n z^n, \quad g(z) = \sum_{n=0}^\infty g_n z^n,
  \]
and can also be expressed as an integral with respect to a two-dimensional positive measure; see \cite{MR3816055} for the latter. We remark that the $p$-Fock space $\mathcal{F}_p$ is a reproducing kernel Hilbert space with reproducing kernel
\[
    k_p(z, \omega) = \sum_{n=0}^\infty \frac{z^n \overline{\omega}^n}{(n!)^p}, \]
since the point-evaluation functionals are continuous on the corresponding Fock space.

Note that the space $\mathcal F_2$ was considered in \cite[Lemma 4, p. 181]{Karp2}, and
plays an important role in the theory of discrete analytic functions, see \cite{ajsv} for details.

In these spaces the backward-shift operator has the following properties:
\begin{theorem}
  Let $p\in \mathbb{N}.$ Then:
    \begin{enumerate}[(1)]
  \item The backward-shift operator $R_0$ is a contraction in $\mathcal{F}_p$.
\item Its adjoint operator $R_0^\ast$ is given by
\[
    R_0^* := (IR_0)^{p-1} I,
\]where $I$ stands for the integration operator (\ref{IntOp}).
\end{enumerate}
\end{theorem}
\begin{proof}  The proof that $R_0$ is bounded in $\mathcal F_p, ~p=2,3, \ldots,$ follows the same lines as the proof in the classic case $p=1.$

For the proof of the second statement, we first observe the action of $(IR_0)^{p-1} I$ on monomials:
$$(IR_0) I z^n = (IR_0) \frac{z^{n+1}}{n+1} = I \left( \frac{z^{n}}{n+1}  \right) = \frac{z^{n+1}}{(n+1)^2}, \quad n=0, 1, 2, \ldots,$$
so that by induction one has
$$(IR_0)^{p-1} I z^n = (IR_0)^{p-1} \frac{z^{n+1}}{n+1} = \frac{z^{n+1}}{(n+1)^p}.$$

Hence, given $f(z) = \sum_{n=0}^\infty f_n {z^n}$ and $g(z) = \sum_{n=0}^\infty g_n {z^n}$ in $\mathcal F_p$ we obtain
\begin{eqnarray}
    \langle f, (IR_0)^{p-1}I g \rangle_{p} & = &  \left\langle f , (IR_0)^{p-1}I \sum_{n=0} g_n z^n \right\rangle_{p} \nonumber \\
  & = &  \left\langle f , \sum_{n=1}^\infty g_{n-1} \frac{z^{n}}{n^p} \right\rangle_{p} \nonumber   \\
   & = &   \sum_{n=1}^\infty f_n g_{n-1}  \frac{(n!)^p}{n^p}  \nonumber \\
     & = &   \sum_{n=0}^\infty f_{n+1} g_{n}  (n!)^p \nonumber   \\
& = & \left\langle  \sum_{n=0}^\infty f_{n+1} z^n, g  \right\rangle_{p}   \nonumber   \\
& = & \langle  R_0 f, g  \rangle_{p}. \label{AdjointR0}
\end{eqnarray}
Therefore, $R_0^* = (IR_0)^{p-1}I$ in the Fock space $\mathcal{F}_p$.
\end{proof}

\begin{remark} In~\cite{MR3816055} it was shown that for the multiplication operator and the derivative operator, i.e. with $A=M_z$ and $B=\partial_z$, we have a similar formula for the adjoint $A^*=(BA)^{p-1}B$ in $\mathcal F_p$.
\end{remark}

Now, we want to discuss the commutation relations in terms of diagonal operators.



By  a diagonal operator we mean a (possibly unbounded) linear operator such that
$$D z^n=d_n z^n,\quad n=0,1,\ldots$$
As usual, we denote such diagonal operator by
\begin{equation} \label{D_operator}
D = {\rm diag}\,(d_0,d_1,d_2, \ldots),
\end{equation}
and we denote by $\mathcal D$ the space of such operators. We remark that $\mathcal D$ is a commutative ring. \\

For given $D\in \mathcal D$ we define its forward and backwards diagonal shifts as
\begin{eqnarray}
  D^{(1)}&=&{\rm diag}\,(0,d_0,d_1,\ldots) \label{d(1)} \\
  D^{(-1)}&=&{\rm diag}\,(d_1,d_2, d_3, \ldots),     \label{d(-1)}
\end{eqnarray} that is to say, $D^{(1)} z^n = d_{n-1}z^n$ (under the convention that $d_{-1}=0$) and $D^{(-1)} z^n = d_{n+1}z^n$.
We note that
\begin{equation}
  \left(  D^{(1)}\right)^{(-1)}=D,\quad{\rm while}\quad \left(  D^{(-1)}\right)^{(1)}=PD,\quad  \forall D\in\mathcal D,
  \end{equation}where $P:={\rm diag}\,(0,1,1,1,\ldots)$.

We also recall
\begin{equation}\label{R_0I}
R_0 I z^n = R_0 \frac{z^{n+1}}{n+1} = \frac{z^{n}}{n+1}, \quad IR_0  z^n = \left\{ \begin{array}{cc}
0, & n=0 \\
I z^{n-1} = \frac{z^n}{n},& n=1,2, \ldots
\end{array}  \right. .
\end{equation}
Although neither $R_0$, neither $I$, belong to $\mathcal D,$ we have $[R_0,I] = D_0 \in\mathcal D$ where $D_0$ is the diagonal operator \eqref{d0} linked to the Fock space $\mathcal F_1$.

%
This allows us to state the following lemma.

\begin{lemma}      \label{lemma4-1}
On the linear span of the polynomials, it holds:
\begin{enumerate}
    \item $DI=ID^{(-1)};$
    \item $ID=D^{(1)}I;$
    \item $DR_0=R_0D^{(1)};$
    \item $D^{(-1)}R_0 = R_0D,$
    \end{enumerate}for every $D\in\mathcal D$.
\end{lemma}

\begin{proof}
Under the usual convention $0 z^{-1}=0$, we have
\begin{enumerate}
    \item $DI z^n = D \frac{z^{n+1}}{n+1} = d_{n+1}\frac{z^{n+1}}{n+1} = I(d_{n+1} z^n) = I D^{(-1)} z^n;$
    \item $ID z^n = I(d_n z^n) = d_n \frac{{z^{n+1}}}{n+1} = D^{(1)}(\frac{{z^{n+1}}}{n+1}) = D^{(1)}I{z^n};$
    \item $DR_0 {z^n} = D z^{n-1} = d_{n-1} z^{n-1} = R_0( d_{n-1} {z^n} ) = R_0 D^{(1)} {z^n};$
    \item $D^{(-1)}R_0 z^n = D^{(-1)} z^{n-1}= d_{n} z^{n-1} = R_0 d_n z^n = R_0 D {z^n},$
\end{enumerate} for $n \in \mathbb N_0.$
\end{proof}

This lemma induces us to consider diagonal operators with forward and backward shifts of order $m$:
$$D^{(m)}:= {\rm diag}\,(\underbrace{0, \ldots, 0}_{m-times}, d_0, d_1, d_2, \ldots), \quad D^{(-m)}:= {\rm diag}\,(d_m, d_{m+1}, d_{m+2}, \ldots), \qquad m \in \mathbb{N}.$$ This corresponds to
$$D^{(m)} z^n = \left\{ \begin{array}{cc}
d_{n-m} z^n & n \geq m \\
0 & n < m
\end{array} \right., \quad D^{(-m)} z^n = d_{n+m} z^n.$$
Since these $m$-shift operators, as well as $R_0I, IR_0,$ are diagonal operators, they commute. Hence, in the same way as in the previous lemma we have:
\begin{gather}
  D^{(m)}R_0 I = R_0 I D^{(m)}, \quad D^{(m)}I R_0 = I R_0 D^{(m)},\\
  D^{(-m)}R_0 I = R_0 I D^{(-m)}, \quad D^{(-m)}I R_0 = I R_0 D^{(-m)}.
                 \end{gather}
%

Based on this, we now can study linear decompositions of powers of these operators where the coefficients belong to $\mathcal D$.

However, we need an auxiliary lemma before stating our main results.

\begin{lemma} \label{Lem:5.0}For every $D \in \mathcal D$ and $n,k \in \mathbb{N_0}$  it holds
\begin{gather}
D^n R_0^k  = R_0^k  (D^{(k)})^n.
                 \end{gather}
\end{lemma}
\begin{proof} For the left-hand side  we have
$$D^n R_0^k z^m = D^n z^{m-k} = (d_{m-k})^n z^{m-k}, \qquad m \geq k,$$
and zero otherwise, while for the right-hand side we obtain
$$R_0^k (D^{(k)})^n z^m = R_0^k (d_{m-k})^n z^{m} = (d_{m-k})^n z^{m-k}, \qquad m \geq k,$$
with again zero otherwise.
\end{proof}

While we state the lemma in a general form of principal importance for us will be the case of $n=1$.

\begin{theorem} For $R_0$ and $I$ acting on the linear span of polynomials, and $D_0$ defined by \eqref{d0}, it holds
\begin{equation}
    (IR_0)^n = \sum_{k=1}^n \Lambda_{k,n} I^k R_0^k,
    \label{IR0n}
    \end{equation}
where  $\Lambda_{k,n}\in\mathcal D$ are given by the recurrence relation
\begin{equation}
\label{formulaGamma}
    \Lambda_{k,n} = \Lambda_{{k-1},{n-1}} + \left(\sum_{l=1}^k D_0^{(l)}\right)\Lambda_{{k},{n-1}},
\end{equation}
with initial values
\begin{equation}
\label{gamma11}
  \Lambda_{n,n} = {\rm diag}\, (1,1,1,\ldots)=\mathbf{id} \mbox{ and } \Lambda_{0,n} = {\rm diag}\, (0,0,0,\ldots), \quad n \in \mathbb{N}.
\end{equation}
and with
\begin{equation}
  \label{init1}
  \Lambda_{k,n} = {\rm diag}\, (0,0,0,\ldots),
\end{equation} for remaining values $k, n$.
\end{theorem}

\begin{proof}
For $n=1$ we have
$$(IR_0)^1 = \Lambda_{1,1} I R_0,$$
which is obviously true as $\Lambda_{1,1} = {\rm diag}\, (1,1,1,\ldots).$




Proceeding inductively, assume that at rank $n$,
\[
    (I R_0)^n = \sum_{k=1}^n \Lambda_{k,n} I^k R_0^k.
\]
Then,
\[
    (I R_0)^{n+1} = (I R_0)^n I R_0 =  \sum_{k=1}^n \Lambda_{k,n} I^k R_0^k I R_0.
\]
Using \eqref{d0d0} we obtain
\begin{align*}
    I^k R_0^k I R_0 &= I^k R_0^{k-1} (I R_0 + D_0) R_0  \\
    &= I^k R_0^{k-1} I R_0^2 + I^k R_0^{k-1} D_0 R_0 \\
    &=  I^k R_0^{k-1} I R_0^2 + D_0^{(1)} I^k R_0^k
\end{align*}
and reiterating we get
\[
    I^k R_0^k I R_0 = I^{k+1} R_0^{k+1} + (\sum_{l=1}^k D_0^{(l)}) I^k R_0^k.,\quad k=1,2,\ldots
\]
Thus, the original formula becomes
\begin{align*}
    \sum_{k=1}^n \Lambda_{k,n} I^k R_0^k I R_0 &= \sum_{k=1}^n \Lambda_{k,n} (I^{k+1} R_0^{k+1} + (\sum_{l=1}^k D_0^{(l)})I^k R_0^k) \\
    &= \sum_{k=1}^{n+1} (\Lambda_{k-1,n} + \Lambda_{k,n}(\sum_{l=1}^k D_0^{(l)}))I^k R_0^k \\
    &= \sum_{k=1}^{n+1} \Lambda_{k,n+1} I^k R_0^k,
\end{align*}
where the $\Lambda_{k,n+1}$ are defined by the formula \eqref{formulaGamma} at rank $n$ for $j,k=1,\ldots, n$ and $\Lambda_{n+1,n+1}$ is the identity diagonal.
Therefore $(I R_0)^{n+1} = \sum_{k=1}^{n+1} \Lambda_{k,n+1} I^k R_0^k$, and the result holds for all integers.
\end{proof}

Note that the same follows for $(R_0 I)^n$, albeit with $D_0^{(l)}$ replaced by $D_0^{(-l)}$. To make it easier to get a clear idea we now illustrate \eqref{IR0n} and \eqref{formulaGamma} for the cases $n=2$ and $n=3$.\smallskip

{\bf Case $n=2$:} We have
\[
  \begin{split}
    (IR_0)^2&=I(IR_0+D_0)R_0\\
    &=I^2R_0^2+ID_0R_0\\
    &=I^2R_0^2+D_0^{(1)}IR_0
    \end{split}
  \]
  so that $\Lambda_{2,2}= \mathbf{id}$ and $\Lambda_{1,2}=D_0^{(1)}$, and      \eqref{formulaGamma} for $n=2$ and $k=1$ becomes
    \[
\Lambda_{1,2}=\Lambda_{0,1}+D_0^{(1)}\Lambda_{1,1},
      \]
which holds in view of \eqref{init1}-\eqref{init1}.\smallskip

      {\bf Case $n=3$:} Iterating \eqref{d0d0} and using the case $n=2$, we now have:
      \[
        \begin{split}
          (IR_0)^3&=(IR_0)^2(IR_0)\\
          &=(I^2R_0^2+D_0^{(1)}IR_0)(IR_0)\\
          &=I^2R_0(R_0I)R_0+D_0^{(1)}(I^2R_0^2+D_0^{(1)}IR_0)\\
          &=I^2R_0(IR_0+D^{(1)})R_0+D_0^{(1)}I^2_0R_0^2+(D_0^{(1)})^2IR_0\\
          &=I^2(IR_0+D_0)R_0^2 +D_0^{(1)}I^2R_0^2+D_0^{(1)}(I^2_0R_0^2+D_0^{(1)}IR_0)\\
          &=I^3R_0^3+(D_0^{(2)}+2D_0^{(1)})I^2R_0^2+D_0^{(1)}IR_0,
\end{split}
        \]
        so that
        \begin{equation}
          \label{values567}
\Lambda_{3,3}=\mathbf{id},\quad \Lambda_{2,3}=2D_0^{(1)}+D_0^{(2)},\quad {\rm and}\quad \Lambda_{1,3}=(D_0^{(1)})^2.
          \end{equation}

          Equation \eqref{formulaGamma} for $n=3$ and $k=1$ and $k=2$ respectively becomes
          \[
            \begin{split}
              \Lambda_{1,3}&=\Lambda_{0,2}+D^{(1)}_0\Lambda_{1,2}\\
              \Lambda_{2,3}&=\Lambda_{1,2}+(D^{(1)}_0+D_0^{(2)})\Lambda_{2,2},
            \end{split}
          \]
          which are verified by \eqref{values567} in view of \eqref{init1}-\eqref{init1}.\\


We now compute the coefficients $\Lambda_{k,n}$ in terms of $D_0$.

\begin{proposition}
For $D_0^{(l)}$ as defined above,
\[
    \Lambda_{k,n} = \sum_{|\alpha| = n-k} \Big[ \prod_{t=1}^k \Big( \sum_{l=1}^t D_0^{(l)} \Big)^{\alpha_t} \Big]
\]
where $|\alpha|:=\alpha_1 + \dots + \alpha_k$ and $\alpha_i$ are non-negative integers.
\label{lemma3-1}
\end{proposition}
\begin{proof}
To prove this relation, we will show that this definition fulfills the recurrence relation for $\Lambda_{k,n}$.  If $k=0$, there are no elements to sum and thus $\Lambda_{0,n} = 0$.  Now, for $n \geq 1$
\begin{align*}
    \Lambda_{n,n} = \sum_{|\alpha| = 0} \Big[ \prod_{t=1}^n \Big( \sum_{l=1}^t D_0^{(l)} \Big)^{\alpha_t} \Big] = 1.
\end{align*}
Thus, this form satisfies the boundary conditions of our recursive relation. It is useful to introduce the notation
\begin{equation}
  C_t=\sum_{l=1}^t D_0^{(l)}.
\end{equation}
Thus, for $k\in\mathbb N$,
\[
C_t^n=\left(\sum_{l=1}^t D_0^{(l)}\right)^n.
\]

  Then
\begin{align*}
    \Lambda_{k+1,n+1} &= \Lambda_{k,n} + \Big(\sum_{l=1}^{k+1} D_0^{(l)} \Big) \Lambda_{k+1,n} \\
    \sum_{|\alpha| = (n-1)-(k-1)} \Big( \prod_{t=1}^{k+1} C_t^{\alpha_t} \Big) &= \Big( \sum_{\alpha_1 + \dots + \alpha_k = n-k} \Big( \prod_{t=1}^k C_t^{\alpha_t} \Big) \Big)\\& \qquad + C_{k+1} \Big( \sum_{\alpha_1 + \dots + \alpha_{k+1} = n-k-1} \Big( \prod_{t=1}^{k+1} C_t^{\alpha_t} \Big) \Big).
\end{align*}
Rewriting the right hand side with explicit terms for all $\alpha_i$ gives the following:
\[
    \sum_{\alpha_1 + \dots + \alpha_k = n-k} C_1^{\alpha_1} \cdots C_k^{\alpha_k} C_{k+1}^0 + \sum_{\alpha_1 + \dots + \alpha_{k+1} = n-k-1} C_1^{\alpha_1} \cdots C_k^{\alpha_k} C_{k+1}^{\alpha_{k+1}+1}.
\]
Note that the first sum can be seen as all terms in which $\alpha_{k+1} = 0$ for $\sum_{t=1}^{k+1} \alpha_t = n-k$, while the second sum can be seen as all terms in which $\alpha_{k+1} > 0$ for $\sum_{t=1}^{k+1} \alpha_t = n-k$.  Thus, the equality proposed above holds and this is a valid representation for the exact form of $\Lambda_{k,n}$.
\end{proof}

\begin{proposition} For $0<k<n$ we have the formula for the entries of the matrices $\Lambda_{k,n}$
\begin{equation}\label{coeff_calculation_lambda}
(\Lambda_{k,n})_{i,j}  =  \left\{ \begin{array}{cl}
0 &\mbox{   if } i\neq j \mbox{  or } i = j=1 \\
{n-k+1 \choose k-1} &\mbox{   if } i = j=2 \\
\Big( \frac{1}{i-1} \Big)^{n-k}  \sum_{|\alpha| = n-k}  
 \prod_{s=1}^{i-2} \Big( \frac{-s}{i-1-s}  \Big)^{\alpha_s} & \mbox{   if } 2<  i = j \leq k+1 \\
\Big( \frac{1}{i-1} \Big)^{n-k}  \sum_{|\alpha| = n-k}  \prod_{s=1}^{k} \Big( \frac{-s}{i-1-s}  \Big)^{\alpha_s} & \mbox{   if }  i = j > k+1.
\end{array} \right. .
\end{equation}
\end{proposition}

\begin{proof} Since we have
$$    \Lambda_{k,n} = \sum_{|\alpha| = n-k} \Big[ \prod_{t=1}^k \Big( \sum_{l=1}^t D_0^{(l)} \Big)^{\alpha_t} \Big] \in \mathcal{D},$$ the non-diagonal entries will be zero. For the diagonal entries $(\Lambda_{k,n})_{i,i}$ straightforward calculations give
$$\Big[ \sum_{l=1}^t D_0^{(l)}\Big]_{i,i} = \left\{ \begin{array}{cl}
0 & \qquad \mbox{   if } \qquad i = 1\\
\frac{1}{i-1} & \qquad \mbox{   if } \qquad 2\leq i \leq t+1\\
-\frac{t}{(i-1)(i-t-1)} & \qquad \mbox{   if } \qquad  i > t+1.
\end{array} \right.  $$ Hence, we have $(\Lambda_{k,n})_{1,1}=0,$ and
$$(\Lambda_{k,n})_{2,2}=  \sum_{|\alpha| = n-k} 1 = {n-k+1 \choose k-1}.$$
For the remaining entries $i>2$ we obtain
\begin{enumerate}
\item if $k\geq i-1$ then
\begin{gather*}
\Big[ \prod_{t=1}^k \Big( \sum_{l=1}^t D_0^{(l)} \Big)^{\alpha_t} \Big]_{i,i} = \Big[  \prod_{t=i-1}^k \Big( \sum_{l=1}^t D_0^{(l)} \Big)^{\alpha_t}  \prod_{t=1}^{i-2}\Big( \sum_{l=1}^t D_0^{(l)} \Big)^{\alpha_t}  \Big]_{i,i} \\
= \Big( \frac{1}{i-1} \Big)^{\alpha_{i-1} +\cdots + \alpha_k} (-1)^{\alpha_{1} +\cdots + \alpha_{i-2} } \frac{1}{(i-1)^{\alpha_{1} +\cdots + \alpha_{i-2}}} \frac{1^{\alpha_1} 2^{\alpha_2} \cdots (i-2)^{\alpha_{i-2}} }{ (i-2)^{\alpha_1} (i-3)^{\alpha_2} \cdots 1^{\alpha_{i-2}} } \\
=  \Big( \frac{1}{i-1} \Big)^{n-k}  \prod_{s=1}^{i-2} \Big( \frac{-s}{i-1-s}  \Big)^{\alpha_s};
\end{gather*}
\item if $k<i-1,$ then
\begin{gather*}
\Big[ \prod_{t=1}^k \Big( \sum_{l=1}^t D_0^{(l)} \Big)^{\alpha_t} \Big]_{i,i} =  (-1)^{\alpha_{1} +\cdots + \alpha_{k} } \frac{1}{(i-1)^{\alpha_{1} +\cdots + \alpha_{k}}} \frac{1^{\alpha_1} 2^{\alpha_2} \cdots k^{\alpha_{k}} }{ k^{\alpha_1} (k-1)^{\alpha_2} \cdots 1^{\alpha_{k}} } \\
=  \Big( \frac{1}{i-1} \Big)^{n-k}  \prod_{s=1}^{k} \Big( \frac{-s}{i-1-s}  \Big)^{\alpha_s}.
\end{gather*}
\end{enumerate}

\end{proof}


For future computations it will be convenient to introduce the shifted version of the $\Lambda_{k,n}$ under the map \eqref{d(-1)}.

\begin{proposition}
  Let $\Gamma_{k,n}=\Lambda_{k,n}^{(-1)}$.
  Then,
  \begin{equation}
    \label{shift6}
    \Gamma_{k,n} = \sum_{|\alpha| = n-k} \Big[ \prod_{t=1}^k \Big( \sum_{l=1}^t D_0^{(l-1)} \Big)^{\alpha_t} \Big] = \sum_{|\alpha| = n-k} \Big[ \prod_{t=1}^k \Big( \sum_{l=0}^{t-1} D_0^{(l)} \Big)^{\alpha_t} \Big].
\end{equation}
Then
\begin{equation}
  \label{r000}
  R_0 \Lambda_{k,n} = \Gamma_{k,n}R_0,
\end{equation}
  and
  \begin{equation}
    \label{shift4}
  \Gamma_{k,n} = \Gamma_{{k-1},{n-1}} + (\sum_{l=1}^k D_0^{(l-1)})\Gamma_{{k},{n-1}}
  \end{equation}
with the same boundary conditions \eqref{gamma11}-\eqref{init1} as for $\Lambda_{k,n}$.
\end{proposition}

\begin{proof}
\eqref{shift6} follows from Proposition \ref{lemma3-1}, while \eqref{r000} follows from \eqref{d(-1)} and \eqref{shift4} is a consequence of \eqref{formulaGamma}.
\end{proof}

\begin{proposition}
In terms of the entries of the matrices $\Gamma_{k,n}$ ($0<k<n$) we have the formula
\begin{equation}\label{coeff_calculation_gamma}
(\Gamma_{k,n})_{i,j} =   \left\{ \begin{array}{cl}
0 &\mbox{   if } i\neq j   \\
{n-k+1 \choose k-1} &\mbox{   if } i = j=1 \\
  \Big( \frac{1}{i} \Big)^{n-k}  \sum_{|\alpha| = n-k}  \prod_{s=1}^{i-1}  \Big( \frac{-s}{i-s}  \Big)^{\alpha_s}     & \mbox{   if } 2 \leq   i = j \leq k \\
  \Big( \frac{1}{i} \Big)^{n-k} \sum_{|\alpha| = n-k}   \prod_{s=1}^{k} \Big( \frac{-s}{i-s}  \Big)^{\alpha_s} & \mbox{   if }  i = j > k.
\end{array} \right. .
\end{equation}
\end{proposition}

\begin{proof} Since we have
$\Gamma_{k,n} =  \sum_{|\alpha| = n-k} \Big[ \prod_{t=1}^k \Big( \sum_{l=0}^{t-1} D_0^{(l)} \Big)^{\alpha_t} \Big] \in \mathcal{D},$ the non-diagonal entries will be zero. For the diagonal entries $(\Gamma_{k,n})_{i,i}$ again straightforward calculations give
$$\Big[ \sum_{l=0}^{t-1} D_0^{(l)}\Big]_{i,i} = \left\{ \begin{array}{cl}
\frac{1}{i} & \qquad \mbox{   if } \qquad 1\leq i \leq t\\
 & \\
\frac{-t}{i(i-t)} & \qquad \mbox{   if } \qquad  i > t.
\end{array} \right.  $$ Hence, we have again  $(\Lambda_{k,n})_{1,1}= {n-k+1 \choose k-1}$  and for the remaining entries $i \geq 2$ we obtain
\begin{enumerate}
\item if $k\geq i$ then
\begin{gather*}
\Big[ \prod_{t=1}^k \Big( \sum_{l=0}^{t-1} D_0^{(l)} \Big)^{\alpha_t} \Big]_{i,i} = \Big[  \prod_{t=i}^k \Big( \sum_{l=0}^{t-1} D_0^{(l)} \Big)^{\alpha_t}  \prod_{t=1}^{i-1}\Big( \sum_{l=0}^{t-1} D_0^{(l)} \Big)^{\alpha_t}  \Big]_{i,i} \\
= \Big( \frac{1}{i} \Big)^{\alpha_{i} +\cdots + \alpha_k}
\frac{(-1)^{\alpha_{1} +\cdots + \alpha_{i-1}} }{i^{\alpha_{1} +\cdots + \alpha_{i-1}}} \frac{1^{\alpha_1} 2^{\alpha_2} \cdots (i-1)^{\alpha_{i-1}} }{ (i-1)^{\alpha_1} (i-2)^{\alpha_2} \cdots 1^{\alpha_{i-1}} } \\
=   \Big( \frac{1}{i} \Big)^{n-k}  \prod_{s=1}^{i-1} \Big( \frac{-s}{i-s}  \Big)^{\alpha_s};
\end{gather*}
\item if $k<i,$ then
\begin{gather*}
\Big[ \prod_{t=1}^k \Big( \sum_{l=0}^{t-1} D_0^{(l)} \Big)^{\alpha_t} \Big]_{i,i} =   \frac{1}{i^{\alpha_{1} +\cdots + \alpha_{k}}} \frac{(-1)^{\alpha_1} (-2)^{\alpha_2} \cdots (-k)^{\alpha_{k}} }{ (i-1)^{\alpha_1} (i-2)^{\alpha_2} \cdots (i-k)^{\alpha_{k}} } \\
=  \Big( \frac{1}{i} \Big)^{n-k}  \prod_{s=1}^{k} \Big( \frac{-s}{i-s}  \Big)^{\alpha_s}.
\end{gather*}
\end{enumerate}

\end{proof}

This considerations allow us to compute the commutator of $R_0$ and its adjoint.
Recall that $D_0$ defined by \eqref{d0} satisfies $[R_0,I]=D_0$. In $\mathcal F_1$ we have $I^*=R_0$ and so $[R_0^*,R_0]=-D_0$ in that space. We now compute this commutator in a general $\mathcal F_p$ space.
\begin{theorem}
For any $\mathcal{F}_p$ as defined above,
\[
  [R_0, \, R_0^*] = D_0^p + \sum_{k=1}^p ( \Lambda_{{k+1},{p+1}}^{(-1)}-\Lambda_{k,p} ) I^k R_0^k.
\]
\end{theorem}
\begin{proof}
The commutator is given by
\[
  \begin{split}
    [R_0, \, R_0^*] &= R_0 R_0^* - R_0^* R_0 \\
    &= R_0 (IR_0)^{p-1} I-(IR_0)^{p-1}IR_0  \\
    &=R_0 (IR_0)^{p-1} I-(IR_0)^p.
    \end{split}
  \]
  Using \eqref{r000} and the expansion \eqref{IR0n} for $(IR_0)^n$, we have:
\begin{align*}
    [R_0, \, R_0^*] &= R_0 \sum_{k=1}^{p-1} \Lambda_{k,{p-1}} I^k R_0^k I -\sum_{k=1}^p \Lambda_{k,p} I^k R_0^k \\
    &= \sum_{k=1}^{p-1} \Gamma_{k,{p-1}} R_0  I^k R_0^k I- \sum_{k=1}^p \Lambda_{k,p} I^k R_0^k.
\end{align*}

Using standard commutator relations, we obtain
\[
    R_0 I^k R_0^k I = I^{k+1}R_0^{k+1} + (\widehat{C_{k+1}} + \widehat{C_{k}})I^kR_0^k + (\widehat{C_k})^2 I^{k-1} R_0^{k-1}
\]
using the notation
\begin{equation}
  \widehat{C_t}=\sum_{l=1}^t D^{(l-1)}.
\end{equation}
  Then
      \[
    \begin{split}
    [R_0, \, R_0^*] &= \sum_{k=1}^{p-1} \Gamma_{k,{p-1}} \Big(I^{k+1}R_0^{k+1} +
    (\widehat{C_{k+1}} + \widehat{C_{k}})I^kR_0^k + (\widehat{C_k})^2I^{k-1}R_0^{k-1} \Big)-\\
    &\hspace{5mm}- \sum_{k=1}^p \Lambda_{k,p} I^k R_0^k .
    \end{split}
\]
Now, analyzing the first term in the first summation,

\begin{align*}
    \sum_{k=1}^p \Gamma_{k,{p-1}}I^{k+1}R_0^{k+1} &= \sum_{k=2}^{p+1} \Gamma_{k-1,p-1}I^{k}R_0^{k} = \sum_{k=1}^p \Gamma_{k-1,p-1} I^k R_0^k
\end{align*}
as $\Gamma_{p,{p-1}} = 0$.  Now, we compute
\[
\Gamma_{1,{n}} = (\Gamma_{0,{n-1}} + (\widehat{C_1})\Gamma_{1,{n-1}}) = (\widehat{C_1})^{n-1},\quad n=1,2,\ldots
\]
The third term  in the first summation is then
\begin{align*}
    \sum_{k=1}^p \Gamma_{k,{p-1}}(\widehat{C_k})^2 I^{k-1} R_0^{k-1} &= \sum_{k=0}^{p-1} \Gamma_{{k+1},{p-1}}(\widehat{C_{k+1}})^2 I^{k} R_0^{k} \\
    &= \sum_{k=1}^{p-1} \Gamma_{{k+1},{p-1}}(\widehat{C_{k+1}})^2 I^{k} R_0^{k} + \Gamma_{1,p-1} (\widehat{C_1})^2 \\
    &= \sum_{k=1}^{p} \Gamma_{{k+1},{p-1}}(\widehat{C_{k+1}})^2 I^{k} R_0^{k} + (\widehat{C_1})^p.
\end{align*}
So, the summation becomes
\begin{align*}
  [R_0, R_0^*] &= \sum_{k=1}^p (                 (\Gamma_{k-1,p-1} + \Gamma_{k,p-1}(\widehat{C_{k+1}} + \widehat{C_{k}}) + \Gamma_{{k+1},{p+1}}(\widehat{C_{k+1}})^2))I^k R_0^k + (\widehat{C_1})^2 -\Lambda_{k,p})\\
    &= \sum_{k=1}^p ((\Gamma_{{k},{p}} + \Gamma_{{k+1},{p}}(\widehat{C_{k+1}})))I^k R_0^k + (\widehat{C_1})^2-\Lambda_{k,p} \\
    &= \sum_{k=1}^p (\Gamma_{{k+1},{p+1}}-\Lambda_{k,p}) I^k R_0^k + (\widehat{C_1})^p \\
\end{align*}
So, since $\widehat{C_1} = \sum_{l=1}^1 D_0^{(l-1)} = D_0$,
\[
    [R_0\, R_0^*] = D_0^p + \sum_{k=1}^p (\Gamma_{{k+1},{p+1}}-\Lambda_{k,p}) I^k R_0^k
\]
\end{proof}

\section{The general case}
\setcounter{equation}{0}
\label{Sec4}
We now work in the space $\mathcal H(\varphi)$ (see Definition \ref{Hvarphi}), where
\begin{equation}
  \langle z^n,z^m\rangle=\varphi_n\delta_{n,m},\quad n=0,1,\ldots
  \end{equation}
We now consider two (possibly unbounded but closed) operators $A$ and $B$ on $\mathcal H(\varphi)$
\begin{eqnarray}
  Az^n &=& \begin{cases}\,0, \quad\hspace{9mm} n=0,\\
          a_nz^{n-1},\quad n\ge 1,\end{cases}\\
           B z^n& =& b_n z^{n+1},\quad n=0,,1\ldots
\end{eqnarray}

\begin{lemma}
  The conclusions of Lemma \ref{lemma4-1} are still valid with $A$ instead of $R_0$ and $B$ instead of $I$, namely:
\begin{enumerate}
    \item $DB=BD^{(-1)}$
    \item $BD=D^{(1)}B$
    \item $DA=AD^{(1)}$
    \item $AD=D^{(-1)}A$
    \end{enumerate}
    \label{lemma4-222}

\end{lemma}

\begin{proof}
  The proofs follow directly the arguments of Lemma \ref{lemma4-1}, and We only give the proof of the last claim. For $n\ge 1$ we have:
  \[
AD(z^n)=A(d_n z^n)=d_na_n z^{n-1}=D^{(-1)}A z^n.
  \]
\end{proof}

The commutator $[A,B]\in\mathcal D$, and will be denoted by $D(a,b)$. We have:
  \begin{equation}
    (D(a,b))_n=\begin{cases}\, a_1b_0,\quad\hspace{20mm} n=0\\
      a_{n+1}b_n-a_nb_{n-1},\quad n\ge 1
      \end{cases}
   \end{equation}
   We will assume that $A$ and $B$ are densely defined and closed.

\begin{proposition}
It holds that
\[
(BA)^n =\sum_{k=1}^n\Lambda_{k,n}B^kA^k
\]
where $\Lambda_{k,n}$ is defined as before, with $D =D(a,b)= [A, B]$.
\end{proposition}

\begin{proof}

\end{proof}

\begin{theorem}
  \label{lab123}
The operators $A$ and $B$ satisfy $A^*= B$ if and only if
\begin{equation}
  \label{a*b}
 \overline{a_{n+1}}\varphi_n=b_n\varphi_{n+1},\quad n=0,1,\ldots
\end{equation}
\end{theorem}

\begin{proof}
  For $m\ge 1$,
  \begin{equation}
    \label{le-feu}
  \begin{split}
    \langle A^*z^n,z^m\rangle&=\langle z^n,Az^m\rangle\\
    &=\langle z^n,a_mz^{m-1}\rangle\\
    &=\overline{a_m}\varphi_{n}\delta_{n,m-1}
  \end{split}
\end{equation}
  while
\[
  \begin{split}
    \langle Bz^n,z^m\rangle&=\langle b_nz^{n+1},z^m\rangle\\
        &=b_n\varphi_{n+1}\delta_{n+1,m}
  \end{split}
\]
For $n=m+1$ we have
\[
  \overline{a_{n+1}}\varphi_n=b_n\varphi_{n+1},
\]
that is, \eqref{a*b}.
\end{proof}

An immediate consequence of the above theorem is the following.

\begin{proposition}
The set of sequences $(a,b)$ satisfying \eqref{a*b} is a real infinite-dimensional vector space.
\end{proposition}

\begin{remark}{\rm We have:
    \begin{itemize}
      \item
  For $A=\partial$ and $B=M_z$ we have $a_{n}=n$ and $b_n=1$ and \eqref{a*b} becomes $(n+1)\varphi_n=\varphi_{n+1}$, that is $\varphi_n=c\cdot n!$ for some $c>0$, and we get the Fock space, as
  expected.

  \item For $A=R_0$ and $B=I$ we have
    $a_n=1$ and $b_n=\frac{1}{n+1}$ and \eqref{a*b} becomes now $\varphi_n=\frac{\varphi_{n+1}}{n+1}$, which leads also the same $\varphi_n$ as above.

   \item  If $A=R_0$ and $B=M_z$ we have $a_n=1$ and $b_n=1$, and
     $\varphi_n=c$. Here, we get the Hardy space, but we are not in the setting of entire functions anymore.
   \end{itemize}
   }
\end{remark}

\begin{remark}
  {\rm Since we assumed the operators to be closed, $A^*=B$ is equivalent to $A=B^*$.}
\end{remark}

\begin{example}
A solution to  \eqref{a*b} is given by
\begin{equation}
  \label{123321}
a_{n}=\frac{\varphi_{n}}{\varphi_{n-1}}\quad and\quad b_n=1,
\end{equation}
i.e.
\begin{equation}
    A=\partial^\varphi\quad and\quad B=M_z
    \end{equation}
Another solution to \eqref{a*b} is given by
\begin{equation}
   \label{123321123}
a_{n}=1\quad and\quad b_n=\frac{\varphi_n}{\varphi_{n+1}},
\end{equation}
i.e.
\end{example}
\begin{equation}
    A=R_0 \quad and\quad B=I^\varphi.
    \end{equation}
In the last case when $b_n=n!$ the operator $I^\varphi$ reduces
to the standard integration operator.\\


\begin{theorem}
Let  $p\in \mathbb{N}$, $p\geq 2$.  Then, the operators $A$ and $B$ defined satisfy $A^*= (BA)^{p-1}B$ if and only if
\[
\overline{a_{n+1}}\varphi_n=b_n^pa_{n+1}^{p-1}\varphi_{n+1},\quad n=0,1,\ldots
\]
\end{theorem}
\begin{proof}
  We already computed $\langle A^*z^n,z^m\rangle$ in the proof of Theorem \ref{lab123}.
  On the other hand, for $n\ge 1$, $BAz^n=B(a_nz^{n-1})=a_nb_{n-1}z^n$.
Thus:
\[
  \begin{split}
    \langle (BA)^{p-1}Bz^n,z^m\rangle&=   b_n \langle (BA)^{p-1}z^{n+1},z^m\rangle\\
    &=b_n(a_{n+1}b_n)^{p-1}\delta_{n+1,m}.
\end{split}
\]
Thus, comparing with \eqref{le-feu}, we have
\[
  \overline{a_m}\varphi_{n}\delta_{n,m-1}=b_n(a_{n+1}b_n)^{p-1}\delta_{n+1,m}.
\]
i.e., with $m=n+1$,
\[
\overline{a_{n+1}}\varphi_n=b_n^pa_{n+1}^{p-1}\varphi_{n+1},\quad n=0,1,\ldots
\]
  \end{proof}

\begin{theorem}
Let  $p\in \mathbb{N}$, $p\geq 2$. Then, the operators $A$ and $B$ satisfy
 $B^* = (AB)^{p-1}A$ if and only if
\begin{equation}
\overline{b_{n-1}}\varphi_{n}=a_n^pb_{n-1}^{p-1}\varphi_{n-1},\quad n=1,2,\ldots
\label{equ123}
\end{equation}
\end{theorem}

\begin{proof}
We have
\[
  \begin{split}
    \langle B^*z^n,z^m\rangle&=\langle z^n,Bz^m\rangle\\
    &=\langle z^n,b_m z^{m+1}\rangle\\
    &=\overline{b_m}\varphi_{m+1}\delta_{n,m+1}.
  \end{split}
\]

On the other hand, $ABz^n=A(b_nz^{n+1})=a_{n+1}b_nz^n$ for $n\in\mathbb N_0$.
Thus for $n\ge 1$,:
\[
  \begin{split}
    \langle (AB)^{p-1}Az^n,z^m\rangle&=a_n\langle (AB)^{p-1}z^{n-1},z^m\rangle\\
    &=\langle a_n(a_nb_{n-1})^{p-1} z^{n-1},z^m\rangle\\
    &=a_n(a_nb_{n-1})^{p-1} \delta_{n-1,m}\varphi_{n-1}.
\end{split}
\]
Thus,
\[
  \overline{b_m}\varphi_{m+1}\delta_{n,m+1}=a_n(a_nb_{n-1})^{p-1} \delta_{n-1,m}\varphi_{n-1}.
  \]
Setting $n=m+1$ we obtain
\[
\overline{b_{n-1}}\varphi_{n}=a_n^pb_{n-1}^{p-1}\varphi_{n-1},\quad n=1,2,\ldots
\]
\end{proof}
Thus, given such an operator $A$, these formulas can be used to find
an associated integration operator such that the adjoint of A can be
expanded in this way. Similarly, given B one can find an associated
differentiation operator for this expansion.\\

In this case, for $A$ and $B$ satisfying $A^* = (BA)^{m-1}B$ and with $D = [A, B]$, we have
\[
[A^*, A] = D^m+\sum_{k=1}^m\left(\Lambda_{k,m}-\Lambda_{{k+1},{m+1}}\right)B^kA^k
\]
and for those satisfying $B^* = (AB)^{m-1}A$, and with $D = [B, A]$, we have
\[
[B^*, B] = D^m +\sum_{k=1}^m\left(\Lambda_{k,m}-\Lambda_{{k+1},{m+1}}\right)A^kB^k.
\]
An important case to consider is when $D$ is the identity. In this case,
$\Lambda_{k,n}=\Gamma_{k,n}=S(n, k)$ and this expansion simplifies to
\[
[B^*, B] = I +\sum_{k=1}^m(k + 1)S(n, k + 1)A^k B^k
\]
reproducing the results from \cite{MR3816055} for $B = M_z$ and $A = \partial_z$. The recurrence relation for these numbers can then be reformatted to yield
\[
(k + 1)S(n, k + 1) = S(n, k)-S(n + 1, k + 1)
\]

Note that, if we take $D$ to the the identity operator, then $\Lambda_{k,n} = \Gamma_{k,n} = S(n, k)$ where $S(n, k)$ are the Stirling numbers of the second kind.  The recurrence relation for these numbers can then be rewritten to yield
\[
    (k+1)S(n, k+1) = S(n, k) - S(n+1, k+1)
\]
and using this in our formula above reproduces the formula from \cite{MR3816055}.

We are now looking at some concrete examples. In the case of $A=R_0$ and $B=I^\varphi$ we have $D=[R_0,I^\varphi]$ is a diagonal operator,
$$D = {\rm diag}\, (d_0, d_1, d_2, \ldots)$$
where the diagonal entries are given by
$$
d_i =   \frac{\varphi_i}{\varphi_{i+1}}-\frac{\varphi_{i-1}}{\varphi_i},\quad i\ge 0,
$$ under the convention $\varphi_{-1}=0.$

This leads to the following expressions of the matrix entries for $\Lambda_{k,n}$ and $\Gamma_{k,n}$.

\begin{proposition} \label{Prop:5.10}
In terms of the entries of both matrices $\Lambda_{k,n}$ and $\Gamma_{k,n}$ ($0<k<n$) we have
\begin{eqnarray}\label{coeff_calculation_lambda_gen}
(\Lambda_{k,n})_{i,j} & = & \left\{  \begin{array}{cc}
 					0 & \mbox{   if }  i\neq j \mbox{   or }  i = j =1 \\
					& \\					
					{n-k+1 \choose k-1} \left( \frac{\varphi_{0}}{\varphi_{1}} \right)^{n-k}, &  \mbox{   if }  2 = i = j \\
					& \\
					\sum_{|\alpha|=n-k} \Big(\frac{\varphi_{i-2}}{\varphi_{i-1}} \Big)^{\alpha_{i-1} +\cdots + \alpha_k}  \prod_{t=1}^{i-2}\Big( \frac{\varphi_{i-2}}{\varphi_{i-1}}- \frac{\varphi_{i-t-2}}{\varphi_{i-t-1}}\Big)^{\alpha_t}, &  \mbox{   if }  2 < i = j \leq k+1 \\
						 & \\
						\sum_{|\alpha|=n-k} \prod_{t=1}^{k}\Big( \frac{\varphi_{i-2}}{\varphi_{i-1}}- \frac{\varphi_{i-t-2}}{\varphi_{i-t-1}}\Big)^{\alpha_t} & \mbox{   if }   i = j > k+1,
						 \end{array} \right.
\end{eqnarray} and
\begin{eqnarray}\label{coeff_calculation_gamma_gen}
(\Gamma_{k,n})_{i,j} & = & \left\{  \begin{array}{cc}					
						 0 & \mbox{   if }  i\neq j \\
					& \\					
					{n-k+1 \choose k-1} \left( \frac{\varphi_{0}}{\varphi_{1}} \right)^{n-k}, &  \mbox{   if }  1 = i = j \\
					& \\
						\sum_{|\alpha|=n-k} \Big(\frac{\varphi_{i-1}}{\varphi_{i}} \Big)^{\alpha_{i} +\cdots + \alpha_k}  \prod_{t=1}^{i-1}\Big( \frac{\varphi_{i-1}}{\varphi_{i}}- \frac{\varphi_{i-t-1}}{\varphi_{i-t}}\Big)^{\alpha_t} & \mbox{   if }  1< i = j \leq k \\
						 & \\
						\sum_{|\alpha|=n-k} \prod_{t=1}^{k}\Big( \frac{\varphi_{i-1}}{\varphi_{i}}- \frac{\varphi_{i-t-1}}{\varphi_{i-t}}\Big)^{\alpha_t} & \mbox{   if }   i = j > k,
						 \end{array} \right..
\end{eqnarray}
\end{proposition}

\begin{proof} The proof follows the same lines as in Propositions 3.8 and 3.9. Again, we have both $ \Lambda_{k,n}, \Gamma_{k,n}  \in \mathcal{D},$ so that all non-diagonal entries are zero. For $(\Lambda_{k,n})_{i,i}$ we have
$$(\Lambda_{k,n})_{i,i}  =  \sum_{|\alpha|=n-k} \left[ \prod_{t=1}^k \Big( \sum_{l=1}^t D^{(l)}\Big)_{i,i} \right],$$
so that $(\Lambda_{k,n})_{1,1}=0.$

For the remaining values straightforward calculations give
$$\Big( \sum_{l=1}^t D^{(l)}\Big)_{i,i} = \left\{  \begin{array}{cc}
						 \frac{\varphi_{i-2}}{\varphi_{i-1}}, &  \mbox{   if }  2\leq i \leq t+1  \\
						 & \\
						\frac{\varphi_{i-2}}{\varphi_{i-1}}- \frac{\varphi_{i-t-2}}{\varphi_{i-t-1}}  & \mbox{   if }  i > t+1
						 \end{array} \right.,
 $$
so that  we obtain for the $\Lambda_{k,n}$-entries
\begin{enumerate}
\item if $i=2$ then
\begin{gather*}
\Big[ \prod_{t=1}^k \Big( \sum_{l=1}^t D^{(l)} \Big)^{\alpha_t} \Big]_{i,i} = \Big(\frac{\varphi_{0}}{\varphi_{1}} \Big)^{n-k};
\end{gather*}
\item if $2< i \leq k+1$ then
\begin{gather*}
\Big[ \prod_{t=1}^k \Big( \sum_{l=1}^t D^{(l)} \Big)^{\alpha_t} \Big]_{i,i} = \Big[  \prod_{t=i-1}^k \Big( \sum_{l=1}^t D^{(l)} \Big)^{\alpha_t}  \prod_{t=1}^{i-2}\Big( \sum_{l=1}^t D^{(l)} \Big)^{\alpha_t}  \Big]_{i,i} \\
= \Big(\frac{\varphi_{i-2}}{\varphi_{i-1}} \Big)^{\alpha_{i-1} +\cdots + \alpha_k}  \prod_{t=1}^{i-2}\Big( \frac{\varphi_{i-2}}{\varphi_{i-1}}- \frac{\varphi_{i-t-2}}{\varphi_{i-t-1}}\Big)^{\alpha_t}  ;
\end{gather*}
\item if $i > k+1,$ then
\begin{gather*}
\Big[ \prod_{t=1}^k \Big( \sum_{l=1}^t D^{(l)} \Big)^{\alpha_t} \Big]_{i,i} =  \prod_{t=1}^{k}\Big( \frac{\varphi_{i-2}}{\varphi_{i-1}}- \frac{\varphi_{i-t-2}}{\varphi_{i-t-1}}\Big)^{\alpha_t}.
\end{gather*}
\end{enumerate} Therefore, the result holds true for the $\Lambda_{k,n}$-entries. In a similar way, we have for the $\Gamma_{k,n}$-entries that
$$\Big( \sum_{l=0}^{t-1} D^{(l)}\Big)_{i,i} = \left\{  \begin{array}{cc}
						 \frac{\varphi_{i-1}}{\varphi_{i}}, &  \mbox{   if }  1\leq i \leq t  \\
						 & \\
						\frac{\varphi_{i-1}}{\varphi_{i}}- \frac{\varphi_{i-t-1}}{\varphi_{i-t}}  & \mbox{   if }  i > t
						 \end{array} \right.,
 $$ so that
\begin{enumerate}
\item if $i=1$ then
\begin{gather*}
\Big[ \prod_{t=1}^k \Big( \sum_{l=0}^{t-1} D^{(l)} \Big)^{\alpha_t} \Big]_{i,i} = \Big(\frac{\varphi_{0}}{\varphi_{1}} \Big)^{n-k};
\end{gather*}
\item if $1< i \leq k$ then
\begin{gather*}
\Big[ \prod_{t=1}^k \Big( \sum_{l=0}^{t-1} D^{(l)} \Big)^{\alpha_t} \Big]_{i,i} = \Big[  \prod_{t=i}^k \Big( \sum_{l=0}^{t-1} D^{(l)} \Big)^{\alpha_t}  \prod_{t=1}^{i-1}\Big( \sum_{l=0}^{t-1} D^{(l)} \Big)^{\alpha_t}  \Big]_{i,i} \\
= \Big(\frac{\varphi_{i-1}}{\varphi_{i}} \Big)^{\alpha_{i} +\cdots + \alpha_k}  \prod_{t=1}^{i-1}\Big( \frac{\varphi_{i-1}}{\varphi_{i}}- \frac{\varphi_{i-t-1}}{\varphi_{i-t}}\Big)^{\alpha_t}  ;
\end{gather*}
\item if $i > k,$ then
\begin{gather*}
\Big[ \prod_{t=1}^k \Big( \sum_{l=0}^{t-1} D^{(l)} \Big)^{\alpha_t} \Big]_{i,i} =  \prod_{t=1}^{k}\Big( \frac{\varphi_{i-1}}{\varphi_{i}}- \frac{\varphi_{i-t-1}}{\varphi_{i-t}}\Big)^{\alpha_t}.
\end{gather*}
\end{enumerate}

\end{proof}

We conclude this section with an application of Proposition \ref{Prop:5.10}. 

\begin{example} We consider the rank-one case of a Dunkl operator as in Example \ref{Example:3.10}, that is linked to the function
$$
\varphi(z)=e^{z} {}_1F_1(\kappa,2\kappa+1;-2z),
$$
and with coefficients
$$\varphi_{2n}= \frac{(2n)!\left(\kappa+\frac{1}{2}\right)_n}{\left(\frac{1}{2}\right)_n}\quad
  \mbox{ and }\quad  \varphi_{2n+1}=\frac{(2n+1)!\left(\kappa+\frac{1}{2}\right)_{n+1}}{\left(\frac{1}{2}\right)_{n+1}}.$$
 As seen before (see (\ref{3.15})), we have
 $$ \frac{\varphi_{2n}}{\varphi_{2n+1}} = \frac{1}{2n+2 \kappa + 1}, \qquad \frac{\varphi_{2n-1}}{\varphi_{2n}} = \frac{1}{2n}.$$
 Therefore, we have for the diagonal entries of $\Lambda_{k,n}$
 \begin{enumerate}
 \item $(\Lambda_{k,n})_{1,1} =0;$
 \item $(\Lambda_{k,n})_{2,2} = {n-k+1 \choose k-1} \left( \frac{1}{1+2\kappa}\right)^{n-k};$
  \item if $2 < i  \leq k+1$ and
  \begin{itemize}
 \item $i$ even, we have
 \begin{gather*}
  (\Lambda_{k,n})_{i,i} = \sum_{|\alpha|=n-k} \Big(\frac{\varphi_{i-2}}{\varphi_{i-1}} \Big)^{\alpha_{i-1} +\cdots + \alpha_k}  \prod_{t=1}^{i-2}\Big( \frac{\varphi_{i-2}}{\varphi_{i-1}}- \frac{\varphi_{i-t-2}}{\varphi_{i-t-1}}\Big)^{\alpha_t} \\
  = \sum_{|\alpha|=n-k} \Big(\frac{1}{i-1+2\kappa} \Big)^{\alpha_{i-1} +\cdots + \alpha_{k}} \prod_{l=1}^{i/2-1}\Big( \frac{1}{i-1+2\kappa} - \frac{1}{i-2l} \Big)^{\alpha_{2l-1}} \\
  \times \prod_{l=1}^{i/2-1}\Big( \frac{1}{i-1+2\kappa} -  \frac{1}{i-1-2l+2\kappa}  \Big)^{\alpha_{2l}}\\
   = \Big(\frac{1}{i-1+2\kappa} \Big)^{\alpha_{n-k} }  \sum_{|\alpha|=n-k} \prod_{l=1}^{i/2 -1}\Big( \frac{1-2l-2\kappa}{ i-2l} \Big)^{\alpha_{2l-1}} \Big( \frac{-2l}{ i-1-2l+2\kappa}  \Big)^{\alpha_{2l}}.
   \end{gather*}
     \item $i$ odd, we have
 \begin{gather*}
  (\Lambda_{k,n})_{i,i} = \sum_{|\alpha|=n-k} \Big(\frac{\varphi_{i-2}}{\varphi_{i-1}} \Big)^{\alpha_{i-1} +\cdots + \alpha_k}  \prod_{t=1}^{i-2}\Big( \frac{\varphi_{i-2}}{\varphi_{i-1}}- \frac{\varphi_{i-t-2}}{\varphi_{i-t-1}}\Big)^{\alpha_t} \\
  = \sum_{|\alpha|=n-k} \Big(\frac{1}{i-1} \Big)^{\alpha_{i-1} +\cdots + \alpha_{k}} \prod_{l=1}^{(i-1)/2}\Big( \frac{1}{i-1} - \frac{1}{i-2l+2\kappa} \Big)^{\alpha_{2l-1}} \\
  \times \prod_{l=1}^{(i-1)/2-1}\Big( \frac{1}{i-1} -  \frac{1}{i-1-2l }  \Big)^{\alpha_{2l}}\\
   = \Big(\frac{1}{i-1} \Big)^{\alpha_{n-k} }  \sum_{|\alpha|=n-k}  \Big( \frac{2-i+2\kappa }{1 +2\kappa} \Big)^{\alpha_{i-2}}  \prod_{l=1}^{(i-1)/2-1}\Big( \frac{1-2l+2\kappa }{i-2l +2\kappa} \Big)^{\alpha_{2l-1}} \Big( \frac{-2l}{i-1-2l }  \Big)^{\alpha_{2l}}.
      \end{gather*}
 \end{itemize}

 \item if $i >k+1$ and
 \begin{itemize}
 \item $i$ even, we have for $k=2m$
 \begin{gather*}
 (\Lambda_{k,n})_{i,i} = \sum_{|\alpha|=n-k} \prod_{t=1}^{k}\Big( \frac{\varphi_{i-2}}{\varphi_{i-1}}- \frac{\varphi_{i-t-2}}{\varphi_{i-t-1}}\Big)^{\alpha_t} \\
 = \sum_{|\alpha|=n-2m} \prod_{l=1}^{m} \left(\frac{1}{i-1+2\kappa} - \frac{1}{i-1-2l +2\kappa} \right)^{\alpha_{2l}} \left(\frac{1}{i-1+2\kappa} - \frac{1}{i-2l} \right)^{\alpha_{2l-1}} \\
= \left(\frac{-1}{i-1+2\kappa} \right)^{n-2m}  \sum_{|\alpha|=n-2m} \prod_{l=1}^{m} \left(\frac{2l}{i-1-2l +2\kappa} \right)^{\alpha_{2l}} \left(\frac{2l+2\kappa-1 }{ i-2l} \right)^{\alpha_{2l-1}},
 \end{gather*} while for $k=2m+1$ we get
  \begin{gather*}
 (\Lambda_{k,n})_{i,i} = \sum_{|\alpha|=n-k} \prod_{t=1}^{k}\Big( \frac{\varphi_{i-2}}{\varphi_{i-1}}- \frac{\varphi_{i-t-2}}{\varphi_{i-t-1}}\Big)^{\alpha_t} \\
 = \sum_{|\alpha|=n-2m-1} \prod_{l=1}^{m} \left(\frac{1}{i-1+2\kappa} - \frac{1}{i-1-2l +2\kappa} \right)^{\alpha_{2l}}  \prod_{l=1}^{m+1} \left(\frac{1}{i-1+2\kappa} - \frac{1}{i-2l} \right)^{\alpha_{2l-1}}  \\
 = \left(\frac{-1}{i-1+2\kappa} \right)^{n-2m-1}  \sum_{|\alpha|=n-2m-1}  \left(\frac{2m+2\kappa+1 }{ i-2m-2} \right)^{\alpha_{2m+1}}  \prod_{l=1}^{m} \left(\frac{2l}{i-1-2l +2\kappa} \right)^{\alpha_{2l}} \left(\frac{2l+2\kappa-1 }{ i-2l} \right)^{\alpha_{2l-1}}
 \end{gather*}
  \item $i$ odd, we have for $k=2m$
 \begin{gather*}
 (\Lambda_{k,n})_{i,i} = \sum_{|\alpha|=n-k} \prod_{t=1}^{k}\Big( \frac{\varphi_{i-2}}{\varphi_{i-1}}- \frac{\varphi_{i-t-2}}{\varphi_{i-t-1}}\Big)^{\alpha_t} \\
 = \sum_{|\alpha|=n-2m} \prod_{l=1}^{m} \left(\frac{1}{i-1} - \frac{1}{i-1-2l} \right)^{\alpha_{2l}} \left(\frac{1}{i-1} - \frac{1}{i-2l+2\kappa} \right)^{\alpha_{2l-1}} \\
= \left(\frac{-1}{i-1} \right)^{n-2m}  \sum_{|\alpha|=n-2m} \prod_{l=1}^{m} \left(\frac{2l}{i-1-2l } \right)^{\alpha_{2l}} \left(\frac{2l-2\kappa-1 }{ i-2l+2\kappa} \right)^{\alpha_{2l-1}},
 \end{gather*} while for $k=2m+1$ we get
  \begin{gather*}
 (\Lambda_{k,n})_{i,i} = \sum_{|\alpha|=n-k} \prod_{t=1}^{k}\Big( \frac{\varphi_{i-2}}{\varphi_{i-1}}- \frac{\varphi_{i-t-2}}{\varphi_{i-t-1}}\Big)^{\alpha_t} \\
 = \sum_{|\alpha|=n-2m-1} \prod_{l=1}^{m} \left(\frac{1}{i-1} - \frac{1}{i-1-2l} \right)^{\alpha_{2l}}  \prod_{l=1}^{m+1} \left(\frac{1}{i-1} - \frac{1}{i-2l +2\kappa} \right)^{\alpha_{2l-1}}  \\
 = \left(\frac{-1}{i-1+2\kappa} \right)^{n-2m-1}  \sum_{|\alpha|=n-2m-1}  \left(\frac{2m-2\kappa+3 }{ i-2m-2 +2\kappa} \right)^{\alpha_{2m+1}}  \prod_{l=1}^{m} \left(\frac{2l}{i-1-2l } \right)^{\alpha_{2l}} \left(\frac{2l-2\kappa-1 }{ i-2l +2\kappa} \right)^{\alpha_{2l-1}}.
 \end{gather*}
 \end{itemize}
\end{enumerate}

\end{example}

\section{Conclusions}
In this paper we provided a general framework on how to handle commutators of operators which act differently according to the basis elements. Using our approach of working with diagonal operators we could obtain representations of the commuting relation both as operators as well as action on the basis elements. This allows a variety of avenues to expand upon these results.
For our concrete formulae the considered operators are those which shift the basis element by one power such as multiplication and backward shift operators. Our method also allows to consider more general shift operators.\smallskip

Another avenue of extension is to study other explicit examples, such as the Hardy space.
Then, $R_0^*=M_z$ and the commutator
\[
[R_0,M_z]f=(R_0M_z-M_zR_0) f =f(0)
\]
corresponding to
\[
D_0={\rm diag}\,(1,0,0,\ldots).
\]
We then leave, as observed earlier, the realm of entire functions.\smallskip

In terms of applications the obtained framework can be used not only to calculate the corresponding Lie algebras, but also to extend methods and results from the case of the standard Fock space to more general situations involving Gelfond-Leontiev derivatives as considered in the paper. In particular, questions like pseudodifferential operators or function theories based on Fischer duality can be considered.

\section*{Acknowledgments}
Daniel Alpay thanks the Foster G. and Mary McGaw Professorship in Mathematical Sciences, which supported this research. He also thanks Professor Alain Yger for introducing him to the work of Ore
Oystein \cite{MR1503119}.

P. Cerejeiras and U. K\"ahler were supported by Portuguese funds through the CIDMA - Center for Research and Development in Mathematics and Applications, and the Portuguese Foundation for Science and Technology (``FCT--Funda\c{c}\~ao para a Ci\^encia e a Tecnologia''), within project UIDB/04106/2020 and UIDP/04106/2020. 

Trevor Kling thanks Schmid College of Science and Technology (Chapman University)
for an undergraduate summer research grant.

\bibliographystyle{plain}
\def\cprime{$'$} \def\cprime{$'$} \def\cprime{$'$}
  \def\lfhook#1{\setbox0=\hbox{#1}{\ooalign{\hidewidth
  \lower1.5ex\hbox{'}\hidewidth\crcr\unhbox0}}} \def\cprime{$'$}
  \def\cprime{$'$} \def\cprime{$'$} \def\cprime{$'$} \def\cprime{$'$}
  \def\cprime{$'$}

\end{document}